\documentclass[aps,prd,onecolumn,preprintnumbers,groupedaddress,showpacs,nofootinbib,amssymb]{revtex4-2}
\usepackage{graphicx,color}
\usepackage{amsmath}
\usepackage{amssymb}
\usepackage{amsfonts}
\usepackage{bm}
\usepackage{cancel}
\usepackage{slashed}
\newcommand{\e}{\mathrm{e}}

\allowdisplaybreaks[4]
\begin{document}

\preprint{KEK-TH-2800, KEK-Cosmo-0407}
\title{Ghost-free non-local $F(R)$ Gravity Compatible with ACT}
\author{Shin'ichi~Nojiri,$^{1,2}$}
\email{nojiri@nagoya-u.jp}
\author{S.D. Odintsov,$^{3,4}$}
\email{odintsov@ice.cat}
\author{V.K. Oikonomou,$^{5,6}$}
\email{v.k.oikonomou1979@gmail.com;voikonomou@gapps.auth.gr}
\affiliation{$^{1)}$ Theory Center, High Energy Accelerator Research Organization (KEK), \\
Oho 1-1, Tsukuba, Ibaraki 305-0801, Japan \\
$^{2)}$ Kobayashi-Maskawa Institute for the Origin of Particles
and the Universe, Nagoya University, Nagoya 464-8602, Japan \\
$^{3)}$ ICREA, Passeig Luis Companys, 23, 08010 Barcelona, Spain\\
$^{4)}$ Institute of Space Sciences (IEEC-CSIC) C. Can Magrans s/n, 08193 Barcelona, Spain\\
$^{5)}$Department of Physics, Aristotle University of
Thessaloniki, Thessaloniki 54124, Greece\\
$^{6)}$ Center for Theoretical Physics, Khazar University, 41
Mehseti Str., Baku, AZ-1096, Azerbaijan}

\begin{abstract}
We confront the ghost-free non-local $F(R)$ gravity theories with
the latest Atacama Cosmology Telescope (ACT) constraints on the
spectral index of the scalar perturbations and the updated
constraints of Planck/BICEP on the tensor-to-scalar ratio. After
reviewing how the ghost-free non-local version of $F(R)$ gravity
can be obtained, we show that the de Sitter solution can be
obtained in this framework. Also, we show that the resulting
theory can be cast in terms of an $F(R,\phi)$ theory of gravity.
We analyze two models of non-local $F(R)$ gravity, one power-law
and the $R^2$ model, and we show that both models can be
compatible with the ACT and updated Planck/BICEP constraints.
\end{abstract}

\maketitle

\section{Introduction}

The latest Dark Energy Spectroscopic Instrument (DESI) data
\cite{DESI:2024uvr} indicated that the dark energy seems to be
dynamical instead of being a constant, and in addition, there is
an apparent possibility that a phantom crossing had occurred in
the recent late-time Universe~\cite{Lee:2025pzo, Ozulker:2025ehg,
Kessler:2025kju, Nojiri:2025low, Vagnozzi:2019ezj}. In addition,
the $\Lambda$-Cold-Dark-Matter model ($\Lambda$CDM) is plagued
with the Hubble tension
problems~\cite{Pedrotti:2024kpn,Jiang:2024xnu,Vagnozzi:2023nrq,Adil:2023exv,Bernui:2023byc,Gariazzo:2021qtg}.
Thus, although General Relativity (GR) seems to pass the local
tests of gravity at an astrophysical level, at the cosmological
level starts to be obsolete, or at least incomplete
phenomenologically. Modified gravity in its various
forms~\cite{Nojiri:2017ncd, Capozziello:2011et,
FaraoniCapozziellobeyondEinstein, Nojiri:2010wj, Odintsov:2023weg}
seems to provide a consistent framework that can easily harbor
phenomena that GR fails to describe. The Occam's razor motivated
modified gravity is $F(R)$ gravity, because the Einstein-Hilbert
action contains a linear power of the Ricci scalar, thus the
simplest generalization of Einstein-Hilbert gravity would be a
function of the Ricci scalar $R$ containing higher powers of the
Ricci scalar. $F(R)$ gravity has been in the mainstream of
modified gravity cosmological and astrophysical phenomenology, see
for example Refs.~\cite{Nojiri:2003ft, Capozziello:2005ku,
Capozziello:2004vh, Capozziello:2018ddp, Hwang:2001pu,
Cognola:2005de, Nojiri:2006gh, Song:2006ej, Capozziello:2008qc,
Bean:2006up, Capozziello:2012ie, Faulkner:2006ub, Olmo:2006eh,
Sawicki:2007tf, Faraoni:2007yn, Carloni:2007yv, Nojiri:2007as,
Capozziello:2007ms, Deruelle:2007pt, Appleby:2008tv,
Dunsby:2010wg, Odintsov:2020nwm, Odintsov:2019mlf,
Odintsov:2019evb, Oikonomou:2020oex, Oikonomou:2020qah,
Huang:2013hsb, Berry:2011pb, Bonanno:2010bt, Gannouji:2008wt,
Oyaizu:2008sr, Oyaizu:2008tb, Brax:2008hh, Cognola:2007zu,
Boehmer:2007glt, Boehmer:2007kx, deSouza:2007zpn, Song:2007da,
Brookfield:2006mq, delaCruz-Dombriz:2006kob, Achitouv:2015yha,
Kopp:2013lea, Sebastiani:2013eqa, Odintsov:2017hbk,
Myrzakulov:2015qaa,
Feng:2022vcx,Pi:2017gih,Wang:2024vfv,Kim:2025dyi} and references
therein.

In addition, non-local theories of gravity are also motivated from
the quantum effective theory perspective~\cite{Modesto:2017sdr,
Belgacem:2017cqo, Koshelev:2016xqb, Wetterich:1997bz} see also
\cite{Capozziello:2025zjt} for a metric affine approach. Moreover,
the recent proposal on the non-local theory of gravity, which also
took into account cosmological phenomena, was given in
\cite{Deser:2007jk}, and further analyzed in
Refs.~\cite{Nojiri:2007uq, ArkaniHamed:2002fu, Nojiri:2010pw,
Joukovskaya:2007nq, Calcagni:2007ef, Jhingan:2008ym,
Capozziello:2008gu, Koshelev:2008ie, Nesseris:2009jf,
Deffayet:2009ca, Calcagni:2009dg, Cognola:2009jx,
Bronnikov:2009az, Calcagni:2010ab, Vernov:2010ui, Barnaby:2010kx,
Dimitrijevic:2019pct, Dialektopoulos:2018iph, Calmet:2018rkj,
Bahamonde:2017sdo, Elizalde:2011su, Bamba:2012ky, Deser:2013uya,
Deser:2019lmm, Maggiore:2014sia, Chan:2012jj, Koshelev:2016xqb,
Zhang:2011uv, Giani:2019xjf, Zhang:2016ykx, Elizalde:2013dlt}.
However, non-local theories of gravity are plagued by ghost
degrees of freedom. In Ref.~\cite{Nojiri:2019dio}, we demonstrated
that non-local $F(R)$ gravity can be cast in a local form by
introducing scalar fields. We showed further that these theories
can be cast in a ghost-free form and also that the de Sitter
vacuum is a solution of these ghost-free theories of gravity. In
this work, we aim to confront the inflationary phenomenology of
ghost-free non-local models of $F(R)$ gravity with the ACT and
updated Planck constraints on the inflationary parameters.
Specifically, the ACT data stunned the scientific community since
these indicated that the spectral index is actually constrained as
follows \cite{ACT:2025fju, ACT:2025tim},
\begin{align}
\label{act}
n_{s}=0.9743 \pm 0.0034,
\end{align}
while the updated Planck/BICEP constraints indicate that the tensor-to-scalar ratio must be \cite{BICEP:2021xfz},
\begin{align}
\label{planck}
r<0.036
\end{align}
at $95\%$ confidence. There is a large stream of studies in the
literature aiming to provide models of gravity that are compatible
with the ACT data~\cite{Kallosh:2025rni, Gao:2025onc, Liu:2025qca,
Yogesh:2025wak, Yi:2025dms, Peng:2025bws, Yin:2025rrs,
Byrnes:2025kit, Wolf:2025ecy, Aoki:2025wld, Gao:2025viy,
Zahoor:2025nuq, Ferreira:2025lrd, Mohammadi:2025gbu,
Choudhury:2025vso, Odintsov:2025wai, Q:2025ycf, Zhu:2025twm,
Kouniatalis:2025orn, Hai:2025wvs, Dioguardi:2025vci,
Yuennan:2025kde, Ajith:2025rvf, Kuralkar:2025hoz, Modak:2025bjv,
Oikonomou:2025xms, Oikonomou:2025htz, Odintsov:2025jky,
Aoki:2025ywt, Ahghari:2025hfy, McDonough:2025lzo,
Chakraborty:2025wqn, NooriGashti:2025gug, Yuennan:2025mlg,
Deb:2025gtk, Afshar:2025ndm, Ellis:2025zrf,
Yuennan:2025tyx,Wang:2025cpp,Qiu:2025uot,Wang:2025dbj,Asaka:2015vza,Choudhury:2025hnu,Singh:2025uyr,Kim:2025dyi}
although one must be cautious with using the ACT data due to
calibration potential issues \cite{Ferreira:2025lrd}. Also in Ref.
\cite{Peng:2026ofs} an AI program has been developed to pin-point
ACT-compatible inflationary scenarios. In this work, we aim to
confront the ghost-free non-local theories of $F(R)$ gravity with
the ACT data and the updated Planck/BICEP constraints. In the next
years, the theory of inflation~\cite{Linde:1981mu,
Rubakov:2017xzr, Linde:2014nna, Lyth:1998xn} will be in the focus
of the stage 4 Cosmic Microwave Background experiments like the
Simons observatory~\cite{SimonsObservatory:2019qwx}, but also in
the focus of future gravitational wave
experiments~\cite{Hild:2010id, Baker:2019nia, Smith:2019wny,
Crowder:2005nr, Smith:2016jqs, Seto:2001qf, Kawamura:2020pcg,
Bull:2018lat, LISACosmologyWorkingGroup:2022jok}, hence analyzing
viable inflationary models is vital for future research. Our
approach in this article is aligned with this line of research.
Note that we will focus on quasi-de Sitter and de Sitter
solutions, since only these can provide a significantly long
inflationary regime, although one can also study power-law
inflationary regimes.

\section{Non-local Gravity Models without Ghost}

Non-local gravity corrections are motivated in cosmology by the
fact that the first quantum corrections of vacuum configuration
scalar field action are the following,
\begin{align}\label{quantumaction}
&\mathcal{S}_{eff}=\int
\mathrm{d}^4x\sqrt{-g}\Big{(}\Lambda_1+\Lambda_2
\mathcal{R}+\Lambda_3\mathcal{R}^2+\Lambda_4 \mathcal{R}_{\mu
\nu}\mathcal{R}^{\mu \nu}+\Lambda_5 \mathcal{R}_{\mu \nu \alpha
\theta}\mathcal{R}^{\mu \nu \alpha \theta}+\Lambda_6 \square
\mathcal{R}\\ \notag &
+\Lambda_7\mathcal{R}\square\mathcal{R}+\Lambda_8 \mathcal{R}_{\mu
\nu}\square \mathcal{R}^{\mu
\nu}+\Lambda_9\mathcal{R}^3+\mathcal{O}(\partial^8)+...\Big{)}\, ,
\end{align}
with the parameters $\Lambda_i$, $i=1,2,...,6$ being dimensionful
constants. Thus, the study of non-local terms is highly motivated
in cosmology. In Ref.~\cite{Nojiri:2019dio}, we developed the
ghost-free and non-local gravity models consistently based on the
$F(R)$ gravity framework. The action of one of the models is given
by,
\begin{align}
\label{nlFRa1}
S= \int d^4 x \sqrt{-g} \left\{ \frac{1}{2\kappa^2}
\left( R - \frac{1}{2} F(R) \Box^{-1} F(R) \right) +
\mathcal{L}_\mathrm{matter} \left( g_{\mu\nu}, \Phi_i \right)
\right\} \, ,
\end{align}
with $\mathcal{L}_\mathrm{matter} \left( g_{\mu\nu}, \Phi_i \right)$ denoting the Lagrangian density of the matter fields $\Phi_i$.
We now review that the model~\eqref{nlFRa1} has no ghost degrees of freedom.
By using a scalar field $\phi$, we rewrite the gravitational action of Eq.~\eqref{nlFRa1} as follows,
\begin{align}
\label{nlFRa2}
S=& \int d^4 x \sqrt{-g} \left\{ \frac{1}{2\kappa^2} \left( R - \frac{1}{2} \partial_\mu \phi \partial^\mu \phi - \phi F(R) \right)
+ \mathcal{L}_\mathrm{matter} \left( g_{\mu\nu}, \Phi_i \right) \right\} \, .
\end{align}
Further, we rewrite the action of Eq.~\eqref{nlFRa2}, by introducing additionally two scalar fields $A$ and $B$, as follows,
\begin{align}
\label{nlFRa3}
S = \int d^4 x \sqrt{-g} \left\{ \frac{1}{2\kappa^2} \left( A
 - \frac{1}{2} \partial_\mu \phi \partial^\mu \phi - \phi F(A)
+ B \left( R - A \right) \right)
+ \mathcal{L}_\mathrm{matter} \left( g_{\mu\nu}, \Phi_i \right)\right\} \, ,
\end{align}
The variation of the action with respect to $A$ gives,
\begin{align}
\label{nlFRa4}
B=1 - \phi F'(A) \, ,
\end{align}
which can be algebraically solved with respect to $A$ as $A=A\left( \phi, B \right)$ in principle.
By substituting the expression of $A$ into the action~\eqref{nlFRa3}, we obtain,
\begin{align}
\label{nlFRa3BB}
S = \int d^4 x \sqrt{-g} \left\{ \frac{1}{2\kappa^2} \left(A\left( \phi, B \right)
 - \frac{1}{2} \partial_\mu \phi \partial^\mu \phi - \phi F\left(A\left( \phi, B \right)\right)
+ B \left( R - A\left( \phi, B \right) \right) \right)
+ \mathcal{L}_\mathrm{matter} \left( g_{\mu\nu}, \Phi_i \right) \right\} \, .
\end{align}
By using the scale transformation,
\begin{align}
\label{nlFRa4BB}
g_{\mu\nu} = \e^{-\sigma} {\tilde g}_{\mu\nu} \, , \quad
\e^\sigma \equiv B\, ,
\end{align}
we obtain the action in the Einstein frame,
\begin{align}
\label{nlFRa5}
S_\mathrm{E} =& \int d^4 x \sqrt{-\tilde g} \left[ \frac{1}{2\kappa^2} \left\{
\tilde R - \frac{3}{2} \partial_\mu \sigma \partial^\mu \sigma
 - \frac{1}{2} \e^{-\sigma} \partial_\mu \phi \partial^\mu \phi
 - U \left( \phi, \sigma \right) \right\}
+ \e^{-2\sigma} \mathcal{L}_\mathrm{matter}
\left( \e^{-\sigma} {\tilde g}_{\mu\nu}, \Phi_i \right) \right] \, , \nonumber \\
U \left( \phi, \sigma \right) \equiv &
\left( - \e^{-2\sigma} + \e^{-\sigma} \right) A\left( \phi , \sigma\right)
 - \phi \e^{-2\sigma} F\left(A \left( \phi , \sigma\right) \right) \, .
\end{align}
The model has no ghosts as long as $\e^\sigma = B=1 - \phi F'(A) > 0$ in \eqref{nlFRa4} by observing the kinetic terms of the scalar fields $\phi$ and $\sigma$ in Eq.~\eqref{nlFRa5}.

\section{Inflationary Phenomenology of Ghost-free non-Local $F(R)$ Gravity Theory}

We now consider the inflationary cosmology by using the model of Eq.~\eqref{nlFRa1}.
During the inflationary regime, the matter fields should play no important role, these could be neglected.
So we start with the following gravitational action,
\begin{align}
\label{modelnewnonlocalghostfreemodel}
S=& \int d^4 x
\sqrt{-g}\frac{1}{2} \left\{ \frac{R}{\kappa^2}
 - \partial_\mu \phi \partial^\mu \phi -2 \phi F(R) \right\} \, .
\end{align}
We now use the following notation,
\begin{align}
\label{ffunction}
f(R,\phi)=\frac{R}{\kappa^2} - \partial_\mu \phi \partial^\mu \phi -2 \phi F(R)\, .
\end{align}
we also define $f_R$ and $f_{\phi}$, as follows,
\begin{align}
\label{frderivative}
f_R=&\, \frac{\partial f(R,\phi)}{\partial R}=\frac{1}{\kappa^2} -2 \phi F'(R)\, , \\
\label{fphiderivative}
f_{\phi}=&\, \frac{\partial f}{\partial \phi}=-2 F(R)\, .
\end{align}
Note that essentially, the gravitational action~\eqref{modelnewnonlocalghostfreemodel} is nothing but a generalized $f(R,\phi)$ gravity model.
By assuming that the background metric is a Friedmann-Lema\^{i}tre-Robertson-Walker (FLRW) Universe with a flat spatial part,
\begin{align}
\label{metricflrw}
ds^2 = - dt^2 + a(t)^2 \sum_{i=1,2,3} \left(dx^i\right)^2\, ,
\end{align}
the variation of the action~\eqref{modelnewnonlocalghostfreemodel} with \eqref{ffunction}, with respect to the metric $g_{\mu\nu}$ and the scalar field $\phi$ leads to the following field equations,
\begin{align}
\label{firstfriedmanequation1}
3H^2=&\, \frac{1}{f_R}\left(\frac{1}{2}\dot{\phi}^2 + \frac{Rf_R-f}{2}-3H\dot{f}_R \right) \, , \\
\label{firstfriedmanequation2}
 -3H^2-2\dot{H}=& \frac{1}{f_R}\left(\frac{1}{2}\dot{\phi}^2 -\frac{Rf_R-f}{2}+\ddot{f}_R+2H\dot{f}_R \right)\, , \\
\label{firstfriedmanequation3}
0 = &\, \ddot{\phi}+3H\dot{\phi}-\frac{1}{2}f_{\phi} \, .
\end{align}
Here, the ``dot'' or ``$\dot\ $'' denotes differentiation with respect to the cosmic time $t$.
For the model~\eqref{modelnewnonlocalghostfreemodel}, the gravitational wave speed is $c_T=1 $ \cite{Hwang:2005hb} and also the sound wave propagation speed of the scalar perturbations is $c_A=1$, since the theory is effectively an $f(R,\phi)$ theory of gravity.
%%%%%%%%%%%%%%%
%%%%%%%%%%%%%%%%%%

In the three equations~\eqref{firstfriedmanequation1}, \eqref{firstfriedmanequation2} and \eqref{firstfriedmanequation3}, only two equations are independent and we use \eqref{firstfriedmanequation1} and \eqref{firstfriedmanequation3} by rewriting them as follows,
\begin{align}
\label{firstfriedmanequation1_BB}
\frac{3}{\kappa^2} H^2=&\left. \left\{ \phi \left( - R F'(R) + F(R) + 6 H F''(R) \dot R + 6 H^2 F'(R) \right) + 6 \dot\phi H F'(R) \right\} \right|_{R=12H^2 + 6\dot H} \, , \\
\label{firstfriedmanequation3_BB}
0 = &\, \ddot{\phi} + 3H\dot{\phi} + F(R)\, ,
\end{align}
By combining Eqs.~\eqref{firstfriedmanequation2}) and
\eqref{firstfriedmanequation3}, we also obtain the following,
\begin{align}
\label{equationofmotionnew}
 -2\dot{H}f_R=\dot{\phi}^2-H\dot{f}_R+\ddot{f}_R\, ,
\end{align}
which we use later.

By solving the above equations~\eqref{firstfriedmanequation1_BB} and \eqref{firstfriedmanequation3_BB}, we can determine the dynamics of the FLRW Universe~\eqref{metricflrw}.

\subsection{de Sitter Spacetime as an Exact Solution}

First, we consider the condition that the model has a solution describing the de Sitter spacetime, where $H$ is a constant, $H=H_0$.
We also assume that $\phi$ is also a constant, $\phi=\phi_0$.
By the assumption, Eqs.~\eqref{firstfriedmanequation1_BB} and \eqref{firstfriedmanequation3_BB} reduce to,
\begin{align}
\label{firstfriedmanequation1_BBB_A}
\frac{3}{\kappa^2} {H_0}^2=&\ \phi_0 \left( - 6{H_0}^2 R F'\left(12{H_0}^2 \right) + F \left( 12 {H_0}^2 \right) \right) \, , \\
\label{firstfriedmanequation3_BBB}
0 = &\, F\left( 12{H_0}^2 \right)\, ,
\end{align}
When Eq.~\eqref{firstfriedmanequation3_BBB} has a positive solution for $12{H_0}^2$, the de Sitter spacetime becomes a solution.
Eq.~\eqref{firstfriedmanequation1_BBB_A} can be solved with respect to $\phi_0$, as follows,
\begin{align}
\label{firstfriedmanequation1_BBBBB} \phi_0 = - \frac{1}{2
\kappa^2 F'\left(12{H_0}^2 \right) }\, ,
\end{align}
and note here that we used Eq.~\eqref{firstfriedmanequation3_BBB}.
Therefore, even in the model~\eqref{nlFRa1}, the de Sitter spacetime can be an exact solution.
We should note that,
\begin{align}
\label{Bcons}
B= 1 - \phi_0 F'\left( 12 {H_0}^2 \right)
= 1 + \frac{1}{2\kappa^2} \, .
\end{align}
Therefore, the ghost-free condition is always satisfied.

\subsection{General Inflationary Dynamics of the Ghost-free $f(R)$ Gravity}

The inflationary dynamics for models of the form $f(R,\phi)$ have been well-studied in Ref.~\cite{Hwang:2005hb}, where the slow-roll indices have been calculated as follows,
\begin{align}
\label{slowrollparameters}
\epsilon_1= %&\,
\frac{\dot{H}}{H^2}\, , \quad
\epsilon_2=\frac{\ddot{\phi}}{H\dot{\phi}}\, ,\quad
\epsilon_3=\frac{\dot{f_R}}{2Hf_R}\, ,\quad
\epsilon_4=\frac{\dot{E}}{2HE}\, , \quad
%\label{functionepsilon}
E\equiv%&\,
\frac{f_R}{\dot{\phi}^2}\left( \dot{\phi^2}+\frac{3{\dot{f}_R}^2}{2f_R}\right)\, .
\end{align}
Then the following spectral index $n_s$ and the tensor-to-scalar ratio $r$ are given by
\begin{align}
\label{spectralindex}
n_s=&\, 1+\frac{2 \left( 2\epsilon_1-\epsilon_2+\epsilon_3-\epsilon_4\right)}{1+\epsilon_1}\, , \\
\label{tensortoscalarration}
r=&\, 16 |\epsilon_1-\epsilon_3|\, .
\end{align}
The inflationary phenomenology of the model \eqref{modelnewnonlocalghostfreemodel} could depend on the functional form of $F(R)$.

We shall assume that the slow-roll condition for the Hubble rate holds during the inflationary era, that is,
\begin{align}
\label{slowrollhubblerate}
\dot{H}\ll H^2\, .
\end{align}
The scalar field is also assumed to satisfy the general slow-roll evolution,
\begin{align}
\label{constnatroll_2}
\ddot{\phi}\ll H\dot{\phi}\, .
\end{align}
We often, however, assume that the scalar field evolves in a constant-roll way, which is a generalization of the slow-roll evolution,
\begin{align}
\label{constantroll}
\ddot{\phi}\sim 3\beta H\dot{\phi}\, .
\end{align}
When $\beta \neq 0$, the scalar field obeys the constant-roll evolution rules \cite{Martin:2012pe, Nojiri:2017qvx}, and when $\beta=0$, the scalar field evolves in the standard slow-roll way.

We now consider if the conditions \eqref{slowrollhubblerate} and \eqref{constnatroll_2} or \eqref{constantroll} can be a solution of Eqs.~\eqref{firstfriedmanequation1_BB} and \eqref{firstfriedmanequation3_BB}.
For this purpose, we expand $H$ and $\phi$ as power series of the cosmological time $t$, as follows,
\begin{align}
\label{expnsn}
H = H_0 + H_1 t + \frac{1}{2} H_2 t^2 + \frac{1}{3!} H_3 t^3 + \mathcal{O} \left( t^4 \right)\, , \quad
\phi = \phi_0 + \phi_1 t + \frac{1}{2} \phi_2 t^2 + \frac{1}{3!} \phi_3 t^3 + \mathcal{O} \left( t^4 \right)\, .
\end{align}
Then the condition~\eqref{constnatroll_2} indicates that,
\begin{align}
\label{slowrollhubblerateexp}
H_1 \ll {H_0}^2 \, , \quad H_2 \ll H_1 H_0 \ll {H_0}^3 \, , \quad H_3 \ll H_0 H_2 \ll {H_0}^4\, , \quad H_3 \ll {H_1}^2 \ll {H_0}^4 \, .
\end{align}
On the other hand, Eq.~\eqref{constantroll} gives,
\begin{align}
\label{constantrollexp}
\phi_2 = 3\beta H_0 \phi_1\, .
\end{align}
By using Eq~\eqref{firstfriedmanequation3_BB}, instead of \eqref{firstfriedmanequation3_BBB}, we obtain,
\begin{align}
\label{firstfriedmanequation3_BB_ld}
0 = \phi_2 + 3H_0 \phi_1 + F\left( R=12 {H_0}^2 + 6 H_1 \right) = 3 H_0 \left(\beta + 1 \right) \phi_1 + F\left( R=12 {H_0}^2 + 6 H_1 \right) \, .
\end{align}
Therefore, we obtain,
\begin{align}
\label{ph12}
\phi_1 = \frac{F\left( R=12 {H_0}^2 + 6 H_1 \right)}{3H_0 \left(\beta + 1 \right) }\, , \quad
\phi_2 = \frac{\beta F\left( R=12 {H_0}^2 + 6 H_1 \right)}{\beta + 1} \, .
\end{align}
On the other hand, Eq.~\eqref{firstfriedmanequation1_BB} gives
\begin{align}
\label{firstfriedmanequation1_BB_1}
\frac{3}{\kappa^2} {H_0}^2=&\, \phi_0 \left( - \left( 12 {H_0}^2 + 6 H_1 \right) F'\left( R=12 {H_0}^2 + 6 H_1 \right) + F\left( R=12 {H_0}^2 + 6 H_1 \right) \right. \nonumber \\
&\, \left. + 6 H_0 F''\left( R=12 {H_0}^2 + 6 H_1 \right) \left( 24 H_0 H_1 + 6 H_2 \right) + 6 {H_0}^2 F'\left( R=12 {H_0}^2 + 6 H_1 \right) \right) \nonumber \\
&\, + 6 \phi_1 H_0 F'\left( R=12 {H_0}^2 + 6 H_1 \right) \, .
\end{align}
We now consider the solution of the above equations.
We should note that $H_0$, $\phi_0$, and $\phi_1$ could be determined by the initial conditions.
By assuming \eqref{slowrollhubblerateexp}, the first equation $H_1 \ll {H_0}^2$ in \eqref{ph12} has the following form,
\begin{align}
\label{ph12B}
\phi_1 \sim \frac{F\left( R=12 {H_0}^2 \right)}{3H_0 \left(\beta + 1 \right)}
+ \frac{2F'\left( R=12 {H_0}^2 \right)}{H_0 \left(\beta + 1 \right)} H_1 \, .
\end{align}
which determines $H_1$ as follows,
\begin{align}
\label{H1}
H_1 \sim \frac{H_0 \left(\beta + 1 \right)}{2F\left( R=12 {H_0}^2 \right)} \left( \phi_1 - \frac{F'\left( R=12 {H_0}^2 \right)}{3H_0 \left(\beta + 1 \right)} \right) \, .
\end{align}
We should note,
\begin{align}
\label{ph1B}
\phi_1 \sim \frac{F\left( R=12 {H_0}^2 \right)}{3H_0 \left(\beta + 1 \right)} \, ,
\end{align}
but the small difference between $\phi_1$ and $\frac{F\left( R=12 {H_0}^2 \right)}{3H_0 \left(\beta + 1 \right)}$ determines $H_1$.

On the other hand, Eq.~\eqref{firstfriedmanequation1_BB_1} can be solved with respect to $H_2$ as follows,
\begin{align}
\label{H2}
H_2 =&\, - 4 H_0 H_1 + \frac{1}{36 H_0 F''\left( R=12 {H_0}^2 + 6 H_1 \right)} \left\{
\frac{3{H_0}^2}{\kappa^2 \phi_0} + \left( 12 {H_0}^2 + 6 H_1 \right) F'\left( R=12 {H_0}^2 + 6 H_1 \right) \right. \nonumber \\
&\, \left. - F\left( R=12 {H_0}^2 + 6 H_1 \right)
 - 6 {H_0}^2 F'\left( R=12 {H_0}^2 + 6 H_1 \right) - \frac{6 \phi_1 H_0 F'\left( R=12 {H_0}^2 + 6 H_1 \right) }{\phi_0} \right\} \, .
\end{align}
$H_1$ in \eqref{H2} could be given by \eqref{H1} and it could be very small but because $H_2 \ll H_1 H_0$ as in the second equation~\eqref{ph12}, we may need more fine tuning in \eqref{H2} and we keep $H_1$.
At leading order of $\frac{H_1}{{H_0}^2}$, we may assume,
\begin{align}
\label{H2B}
0 \sim&\, \frac{3{H_0}^2}{\kappa^2 \phi_0} + 6 {H_0}^2 F'\left( R=12 {H_0}^2 \right)
 - F\left( R=12 {H_0}^2 \right) - \frac{6 \phi_1 H_0 F'\left( R=12 {H_0}^2 \right)}{\phi_0} \, ,
\end{align}
which is a relation between $H_0$ and $\phi_0$ and we may solve \eqref{H2B} with respect to $\phi_0$ as follows,
\begin{align}
\label{phi0}
\phi_0 \sim \frac{6 \phi_1 H_0 F'\left( R=12 {H_0}^2 \right) - \frac{3{H_0}^2}{\kappa^2}}
{6 {H_0}^2 F'\left( R=12 {H_0}^2 \right) - F\left( R=12 {H_0}^2 \right)}
\sim& \frac{ \frac{2F\left( R=12 {H_0}^2 \right) F'\left( R=12 {H_0}^2 \right)}{\beta + 1} - \frac{3{H_0}^2}{\kappa^2}}
{6 {H_0}^2 F'\left( R=12 {H_0}^2 \right) - F\left( R=12 {H_0}^2 \right)} \, .
\end{align}
In the second line, we used Eq.~\eqref{ph1B}.
Since at leading (zeroth) order of $\frac{H_1}{{H_0}^2}$, we have,
\begin{align}
\label{fRs}
f_R=&\, \frac{1}{\kappa^2} -2 \phi F'(R) \sim \frac{1}{\kappa^2} -2 \phi_0 F' \left( 12 {H_0}^2 \right) \nonumber \\
\sim&\, \frac{1}{\kappa^2} - \frac{2 \left\{ \frac{2F\left(12 {H_0}^2\right) F'\left(12 {H_0}^2\right)}{\beta + 1} - \frac{3{H_0}^2}{\kappa^2} \right\} F' \left( 12 {H_0}^2 \right) }
{6 {H_0}^2 F'\left(12 {H_0}^2\right) - F\left(12 {H_0}^2\right)} \, , \nonumber \\
{\dot f}_R =&\, -2 \dot\phi F'(R) - 2 \phi F'' (R) \dot R
= -2 \dot\phi F' \left( 12 H^2 + 6 \dot H \right) -2 \phi F''\left( 12 H^2 + 6 \dot H \right) \left( 24 H \dot H + 6\ddot H \right) \nonumber \\
\sim&\, -2 \phi_1 F' \left( 12 {H_0}^2 \right) - 48 \phi_0 F''\left( 12 {H_0}^2 \right) H_0 H_1 \nonumber \\
\sim&\, -\frac{2 F \left( 12 {H_0}^2 \right) F' \left( 12 {H_0}^2 \right)}{3H_0 \left( \beta + 1 \right)}
 - \frac{ 48 \left\{\frac{2F\left( R=12 {H_0}^2 \right) F'\left( R=12 {H_0}^2 \right)}{\beta + 1} - \frac{3{H_0}^2}{\kappa^2}\right\} F''\left( 12 {H_0}^2 \right) H_0 H_1 }
{6 {H_0}^2 F'\left(12 {H_0}^2\right) - F\left(12 {H_0}^2\right)} \nonumber \\
\sim&\, -\frac{2 F \left( 12 {H_0}^2 \right) F' \left( 12 {H_0}^2 \right)}{3H_0 \left( \beta + 1 \right)} \, , \nonumber \\
{\ddot f}_R =&\, -2 \ddot\phi F' \left( 12 H^2 + 6 \dot H \right) -2 \dot \phi F''\left( 12 H^2 + 6 \dot H \right) \left( 24 H \dot H + 6\ddot H \right)
 -2 \phi F'''\left( 12 H^2 + 6 \dot H \right) \left( 24 H \dot H + 6\ddot H \right)^2 \nonumber \\
&\, -2 F''\left( 12 H^2 + 6 \dot H \right) \left( 24 {\dot H}^2 + 24 H \ddot H + 6\dddot H \right) \nonumber \\
\sim&\, -2 \phi_2 F' \left( 12 {H_0}^2 \right) -48 \phi_1 F''\left( 12 {H_0}^2 \right) H_0 H_1 -1152 \phi_0 F'''\left( 12 {H_0}^2 \right) {H_0}^2 {H_1}^2
-48 \phi_0 F''\left( 12 {H_0}^2 \right) {H_1}^2 \nonumber \\
\sim&\, - \frac{2\beta F\left(12 {H_0}^2\right)F' \left( 12 {H_0}^2 \right) }{\beta + 1}
 - \frac{48F\left(12 {H_0}^2\right) F''\left( 12 {H_0}^2 \right) H_0 H_1}{3H_0 \left(\beta + 1 \right)} \nonumber \\
&\, - \frac{\left(1152 \phi_0 F'''\left( 12 {H_0}^2 \right) {H_0}^2 {H_1}^2 + 48 \phi_0 F''\left( 12 {H_0}^2 \right) {H_1}^2 \right)
\left\{\frac{2F\left( R=12 {H_0}^2 \right) F'\left( R=12 {H_0}^2 \right)}{\beta + 1} - \frac{3{H_0}^2}{\kappa^2}\right\}}
{6 {H_0}^2 F'\left( R=12 {H_0}^2 \right) - F\left( R=12 {H_0}^2 \right)} \nonumber \\
\sim&\, - \frac{2\beta F\left(12 {H_0}^2\right)F' \left( 12 {H_0}^2 \right) }{\beta + 1} \, ,\nonumber \\
E=&\, \frac{f_R}{\dot{\phi}^2}\left( \dot{\phi^2}+\frac{3{\dot{f}_R}^2}{2f_R}\right) \nonumber \\
\sim&\, \frac{1}{\kappa^2} -2 \phi_0 F' \left( 12 {H_0}^2 \right) - \frac{3}{\phi_1} F' \left( 12 {H_0}^2 \right) - \frac{72 \phi_0}{{\phi_1}^2} F''\left( 12 {H_0}^2 \right) H_0 H_1 \nonumber \\
\sim&\, \frac{1}{\kappa^2}
 - \frac{2 \left( \frac{2F\left(12 {H_0}^2\right) F'\left(12 {H_0}^2\right)}{\beta + 1} - \frac{3{H_0}^2}{\kappa^2}\right) F'\left( 12 {H_0}^2 \right)}
{6 {H_0}^2 F'\left(12 {H_0}^2\right) - F\left(12 {H_0}^2\right)}
+ \frac{3}{2}\frac{4 F \left( 12 {H_0}^2 \right)^2 F' \left( 12 {H_0}^2 \right)^2}{9{H_0}^2 \left( \beta + 1 \right)^2}
\frac{9{H_0}^2 \left( \beta + 1 \right)^2}{F \left( 12 {H_0}^2 \right)^2}
\nonumber \\
\sim&\, \frac{1}{\kappa^2}
 - \frac{2 \left( \frac{2F\left(12 {H_0}^2\right) F'\left(12 {H_0}^2\right)}{\beta + 1} - \frac{3{H_0}^2}{\kappa^2}\right) F'\left( 12 {H_0}^2 \right)}
{6 {H_0}^2 F'\left(12 {H_0}^2\right) - F\left(12 {H_0}^2\right)}
+ 6 F' \left( 12 {H_0}^2 \right)^2 \, , \nonumber \\
\dot E =&\, {\dot f}_R + \frac{3 {\dot f}_R {\ddot f}_R}{{\dot\phi}^2} - \frac{3{{\dot f}_R}^2 \ddot \phi}{{\dot\phi}^3} \, , \nonumber \\
\sim &\,
-\frac{2 F \left( 12 {H_0}^2 \right) F' \left( 12 {H_0}^2 \right)}{3H_0 \left( \beta + 1 \right)} \left[ 1
 - 3 \frac{2\beta F\left(12 {H_0}^2\right)F' \left( 12 {H_0}^2 \right) }{\beta + 1}\frac{9{H_0}^2 \left( \beta + 1 \right)^2}{F \left( 12 {H_0}^2 \right)^2} \right.  \nonumber \\
&\, \left. + 3 \frac{ 2 F \left( 12 {H_0}^2 \right) F' \left( 12 {H_0}^2 \right)}{3 H_0 \left( \beta + 1 \right)}
\frac{\beta F \left( 12 {H_0}^2 \right)}{\beta + 1}\frac{27{H_0}^3 \left( \beta + 1 \right)^3}{F \left( 12 {H_0}^2 \right)^3}
\right] \nonumber \\
= &\, -\frac{2 F \left( 12 {H_0}^2 \right) F' \left( 12 {H_0}^2 \right)}{3H_0 \left( \beta + 1 \right)} \left[ 1
 - \frac{54\beta \left( \beta + 1 \right) {H_0}^2 F' \left( 12 {H_0}^2 \right) }{F \left( 12 {H_0}^2 \right)}
+ \frac{54 \beta \left( \beta + 1 \right) {H_0}^2 F' \left( 12 {H_0}^2 \right)}{F \left( 12 {H_0}^2 \right)}\right] \nonumber \\
= &\, -\frac{2 F \left( 12 {H_0}^2 \right) F' \left( 12 {H_0}^2 \right)}{3H_0 \left( \beta + 1 \right)} \, ,
\end{align}
the slow-roll indices in \eqref{slowrollparameters} have the following expressions,
\begin{align}
\label{slowrollparametersFR}
\epsilon_1 \sim &\, \frac{H_1}{{H_0}^2} \, , \nonumber \\
\epsilon_2 \sim &\, \frac{\phi_2}{H_0\phi_1} = 3\beta \, ,\nonumber \\
\epsilon_3 =&\, \frac{\dot{f_R}}{2Hf_R} \nonumber \\
\sim &\, \frac{ -2 \phi_1 F' \left( 12 {H_0}^2 \right) - 48 \phi_0 F''\left( 12 {H_0}^2 \right) H_0 H_1 }
{2 H_0 \left\{ \frac{1}{\kappa^2} -2 \phi_0 F' \left( 12 {H_0}^2 \right) \right\} } \nonumber \\
\sim &\, - \frac{F \left( 12 {H_0}^2 \right) F' \left( 12 {H_0}^2 \right)}{3{H_0}^2 \left( \beta + 1 \right)
\left\{ \frac{1}{\kappa^2} - \frac{2 \left\{ \frac{2F\left(12 {H_0}^2\right) F'\left(12 {H_0}^2\right)}{\beta + 1} - \frac{3{H_0}^2}{\kappa^2} \right\} F' \left( 12 {H_0}^2 \right)}
{6 {H_0}^2 F'\left(12 {H_0}^2\right) - F\left(12 {H_0}^2\right)}\right\}} \, , \nonumber \\
\epsilon_4 \sim &\, \frac{\dot{E}}{2HE}
\sim - \frac{-\frac{2 F \left( 12 {H_0}^2 \right) F' \left( 12 {H_0}^2 \right)}{3H_0 \left( \beta + 1 \right)} }{2H_0 \left\{
\frac{1}{\kappa^2}
 - \frac{2 \left( \frac{2F\left(12 {H_0}^2\right) F'\left(12 {H_0}^2\right)}{\beta + 1} - \frac{3{H_0}^2}{\kappa^2}\right) F'\left( 12 {H_0}^2 \right)}
{6 {H_0}^2 F'\left(12 {H_0}^2\right) - F\left(12 {H_0}^2\right)}
+ 6 F' \left( 12 {H_0}^2 \right)^2 \right\}}
\, .
\end{align}
When we consider the spectral index $n_s$ and the tensor-to-scalar ratio $r$ in \eqref{spectralindex} at leading (zeroth) order of $\frac{H_1}{{H_0}^2}$, we may neglect $\epsilon_1$ and we obtain,
\begin{align}
\label{spectralindexB}
n_s\sim&\, 1 + 2 \left( -\epsilon_2+\epsilon_3-\epsilon_4\right) \, , \\
\label{tensortoscalarrationB}
r\sim&\, 16 |\epsilon_1-\epsilon_3|\, .
\end{align}
We use the above formula by considering the explicit forms of $F(R)$.

\section{Analytic Examples of ACT-Compatible Non-local $F(R)$ Gravity}

In this section, we shall consider two models of non-local $F(R)$ gravity in the form provided in the previous section, which lead to ACT-compatible inflationary phenomenology.
We shall adopt standard slow-roll approximations and analyze the phenomenology of the resulting models in detail.

\subsection{A Power-law $F(R)$ Model}

\begin{figure}
\centering
\includegraphics[width=28pc]{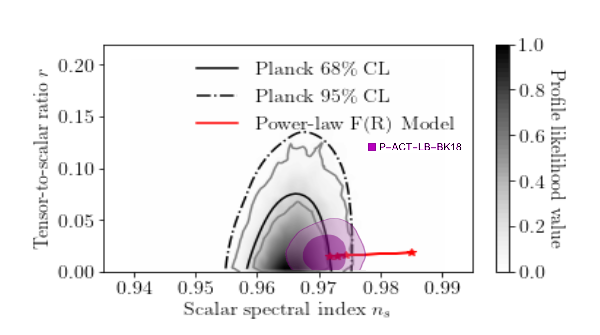}
\caption{Marginalized curves of the Planck 2018 data and the
non-local power-law $F(R)$ gravity model, confronted with the ACT
data, the Planck 2018 data, and the updated Planck constraints on
the tensor-to-scalar ratio for $n=1.7657$, $\alpha=2$, $N=50$ and
$\beta$ in the range $\beta=[0.625,0.626]$. }\label{plot1}
\end{figure}

We first consider the case that the $F(R)$ gravity of Eq.~\eqref{modelnewnonlocalghostfreemodel} has the following general power-law form,
\begin{align}
\label{powerlawfr}
F(R)=-\alpha R^n\, ,
\end{align}
with $n$ being any number in the range $1<n<2$ except for the case $n \neq 2$, which will be considered separately.
Also, $\alpha$ is a dimensionful parameter. Using the constant-roll condition for the scalar field \eqref{constantroll}, the field equation for the scalar field reads,
\begin{align}
\label{eqnmotionforscalarnewpowerlaw} 3H(\beta+1)\dot{\phi}+\alpha R^n=0\, ,
\end{align}
and since the Ricci scalar for the FRW metric is $R=12H^2+6\dot{H}$, taking into account that the slow-roll condition~\eqref{slowrollhubblerate} holds, the Ricci scalar can be approximated as $R\sim 12H^2$ during inflation, hence we can solve \eqref{eqnmotionforscalarnewpowerlaw} with respect to $\dot{\phi}$ and we get,
\begin{align}
\label{solutonfordotphi}
\dot{\phi}\simeq \frac{\gamma H^{2n-1}}{\beta+1}\, ,
\end{align}
with $\gamma$ standing for,
\begin{align}
\label{gammadefinitiolocal}
\gamma=\frac{12^n\alpha}{3}\, .
\end{align}
Thus, for the power-law $F(R)$ model at hand, we have approximately,
\begin{align}
\label{frfunctionderivativepowerlaw}
F_R=\frac{1}{\kappa^2}+2n\phi \alpha R^{n-1}\, ,
\end{align}
and since during the inflationary era, the second term in the above equation overwhelms the evolution (recall $n>1$), we get approximately,
\begin{align}
\label{approximaderivativefR}
f_R\sim 2n \phi \alpha R^{n-1}\, .
\end{align}
Hence, the first Friedman equation, namely, Eq.~\eqref{firstfriedmanequation1}, becomes,
\begin{align}
\label{firstfriednmannequationpowerlaw}
6H^2n\phi \alpha R^{n-1}\simeq \frac{\alpha^2R^{2n}}{2\cdot 3^2 H^2
(\beta+1)^2}+\phi \alpha (n-1)R^n-3H\left( 2n\dot{\phi}\alpha
R^{n-1}+2n (n-1)\phi \alpha R^{n-2}\dot{R}\right)\, .
\end{align}
The last term in the equation above, namely, $\sim \dot{R}$, is subdominant, hence at leading order we omit it.
Therefore, we have,
\begin{align}
\label{resultingequationforphipowerlaw}
\phi \left(6H^2\alpha (-n+2) \right)\simeq \frac{\alpha^2R^{n+1}}{2\cdot 3^2 H^2
(\beta+1)^2}+\frac{2n\alpha^2 R^n}{\beta+1}\, ,
\end{align}
and by using the $R\sim 12H^2$, we obtain,
\begin{align}
\label{dagonia} \phi \sim -A H^{2n-2}\, ,
\end{align}
with $A$ being,
\begin{align}
\label{alphaparameter}
A=\frac{1}{6\alpha (-2+n)}\left(\frac{12^{n+1} \alpha^2} {2\cdot 3^2(\beta+1)^2}+\frac{2\left(12\right)^n n\alpha^2}{\beta+1} \right)\, .
\end{align}
Using Eqs.~\eqref{dagonia} and \eqref{solutonfordotphi}, we can express the slow-roll indices of Eq.~\eqref{slowrollparameters} in terms of the Hubble rate, having in mind the slow-roll conditions~\eqref{slowrollhubblerate}, thus we get the following expressions after some algebra,
\begin{align}
\label{slowrollparameterspowerlawmodel}
\epsilon_1=&\, \frac{\dot{H}}{H^2}\, , \quad \epsilon_2=3\beta\, , \quad
\epsilon_3\simeq \frac{2^{2 n-1} 3^{n-1} \alpha}{A (\beta +1)}+\frac{n \dot{H}}{H^2}-\frac{\dot{H}}{H^2}\, , \nonumber \\
\epsilon_4\simeq&\, -\frac{12 \alpha  n^2 \dot{H}}{H^2 (A-6 \alpha n)} +\frac{2 A n \dot{H}}{H^2 (A-6 \alpha  n)}
+\frac{12 \alpha  n \dot{H}}{H^2 (A-6 \alpha n)}-\frac{2 A \dot{H}}{H^2 (A-6 \alpha n)} \nonumber \\
&\, -\frac{2^{4-2 n} 3^{2-n} n^2 \dot{H}^2 A }{H^4 (A-6 \alpha n)}
+\frac{2^{3-2 n} 3^{3-n} n^2 \dot{H}^2 A}{H^4 (A-6 \alpha n)}
 -\frac{2^{3-2 n} 3^{3-n} n \dot{H}^2 A}{H^4 (A-6 \alpha  n)} \nonumber \\
&\, -\frac{2^{4-2 n} 3^{2-n} n^3 \dot{H}^2 A}{H^4 (A-6 \alpha n)}
+\frac{2^{4-2 n} 3^{2-n} n^2 \dot{H}^2 A \beta}{H^4 (A-6 \alpha n)}
 -\frac{2^{3-2 n} 3^{3-n} n \dot{H}^2 A \beta}{H^4 (A-6 \alpha n)} \nonumber \\
&\, +\frac{2^{3-2 n} 3^{3-n} n^2 \dot{H}^2 A \beta}{H^4 (A-6 \alpha n)} -\frac{2^{4-2 n} 3^{2-n} n^3 \dot{H}^2 A \beta}{H^4 (A-6 \alpha  n)}\, .
\end{align}
Notice that the slow-roll indices $\epsilon_3$ and $\epsilon_4$ can be cast in terms of the slow-roll index $\epsilon_1$, hence we have,
\begin{align}
\label{slowrollparameterspowerlawmodellasttwoindices}
\epsilon_3\simeq&\, \frac{2^{2 n-1} 3^{n-1} \alpha}{A (\beta +1)}+n \epsilon_1-\epsilon_1\, , \nonumber \\
\epsilon_4\simeq&\, -\frac{12 \alpha  n^2 \epsilon_1}{A-6 \alpha n}
+\frac{2 A n \epsilon_1}{A-6 \alpha  n} +\frac{12 \alpha n \epsilon_1}{A-6 \alpha  n}
 -\frac{2 A \epsilon_1}{A-6 \alpha n} \nonumber \\
&\, -\frac{2^{4-2 n} 3^{2-n} n^3 \epsilon_1^2 A}{A-6 \alpha n}
+\frac{2^{4-2n} 3^{2-n} n^2 \epsilon_1^2 A}{A-6 \alpha n}
+\frac{2^{3-2 n} 3^{3-n} n^2 \epsilon_1^2 A}{A-6 \alpha n}
 -\frac{2^{3-2 n} 3^{3-n} n \epsilon_1^2 A}{A-6 \alpha n} \nonumber \\
&\, -\frac{2^{3-2 n} 3^{3-n} n \epsilon_1^2 A \beta}{A-6 \alpha n}
 -\frac{2^{4-2 n} 3^{2-n} n^3 \epsilon_1^2 A \beta}{A-6 \alpha n}
+\frac{2^{4-2 n} 3^{2-n} n^2 \epsilon_1^2 A \beta}{A-6 \alpha n}
+\frac{2^{3-2 n} 3^{3-n} n^2 \epsilon_1^2 A \beta}{A-6 \alpha  n} \, .
\end{align}
Now we must find the Hubble rate as a function of the cosmic time.
This is possible in the power-law $F(R)$ gravity at hand.
This is standard in power-law $F(R)$ gravity, using a standard parametrization.
However, this parametrization leads to inconsistencies in standard power-law $F(R)$ gravity, as it was shown in Ref.~\cite{Odintsov:2025eiv}, nevertheless this is the only method applicable in this $F(R,\phi)$ context, so we shall employ it, having in mind the problematic issues that it may lead to in standard power-law $F(R)$ gravity.
So to proceed, by substituting the evolution for the scalar field $\phi$ and $\dot{\phi}$ from Eqs.~\eqref{dagonia} and \eqref{solutonfordotphi} in the field equation, we get a differential equation for the Hubble rate, and thus by solving it, one is able to find the cosmological evolution during the inflationary era.
At leading order, upon combining Eqs.~\eqref{dagonia} and \eqref{solutonfordotphi}, the differential equation yields the following leading order result,
\begin{align}
\label{diffeqnfriedmanneqn1}
 -4\dot{H}\phi n \alpha R^{n-1}\simeq
 \frac{\gamma^2R^{2n-1}}{12^{n-1}(\beta+1)^2}-H2n\dot{\phi}\alpha R^{n-1}+2n\dot{\phi} \alpha (n-1)R^{n-2}\dot{R}\, ,
\end{align}
and again, by using the approximation $R\sim 12H^2$,  we obtain the following analytic solution,
\begin{align}
\label{hubblesolution}
H(t)=\frac{\alpha  (\beta +1) 2^{2 n} 3^n n (A (-\beta )-A-\gamma +\gamma  n)}{3 \gamma^2 t}\, ,
\end{align}
with $A$ and $\gamma$ being defined in Eqs.~\eqref{alphaparameter} and \eqref{gammadefinitiolocal}.
Now the evolution \eqref{hubblesolution} is a power-law evolution $a(t)\sim t^{\beta}$ but it does not lead to an inflationary evolution, unless $n$ is appropriately constrained in the range $\frac{\sqrt{3}+1}{2}<n<2$.
Now we can proceed to the phenomenology of the model so we can easily extract analytic forms of them, since $\epsilon_1$ is easily derivable, given the Hubble rate of Eq.~\eqref{hubblesolution}.
Indeed, substituting the Hubble rate in $\epsilon_1$, we easily get,
\begin{align}
\label{hubbleslowrollone}
\epsilon_1=-\frac{\gamma^2 2^{-2 n} 3^{1-n}}{\alpha  (\beta +1) n (A (-\beta )-A+\gamma -\gamma  n)}\, ,
\end{align}
and in the same vein, we can easily obtain the observational indices $n_s$ and $r$ appearing in Eqs.~\eqref{spectralindex} and \eqref{tensortoscalarration}.
We quote only the tensor-to-scalar ratio, since the scalar spectral index is too lengthy to be quoted here,
\begin{align}
\label{tensortoscalarpowerlaw}
r=-\frac{24 (n-2) \left(3 (\beta +1)^2 n^3-6 \beta  (\beta +1) n^2+\left(6 \beta^2+\beta -3\right)
n-4\right)}{n (3 (\beta +1) n+1) \left(6 \beta +3 (\beta +1) n^2-6 (\beta +1) n+7\right)}\, .
\end{align}
Note that the resulting observational indices are
$\alpha$-independent, and in addition, these do not depend on the
$e$-foldings number. Thus, this model is plagued with the graceful
exit from inflation issue, at least at the classical level. To
proceed, we found various sets of values of the free parameters
that may provide an ACT-compatible phenomenology, for example, if
we choose Planck units ($\kappa=1$) and take $n=1.7657$,
$\alpha=2$, $\beta=0.6258$, and $N=50$, we get $n_s=0.97438$ and
$r=0.01603$ which are well within the updated Planck constraint
\eqref{planck} and the ACT constraint \eqref{act}. Now, in
Fig.~\ref{plot1} we present the confrontation of the non-local
power-law model with the Planck 2018 data and the updated ACT
data, choosing $n=1.7657$, $\alpha=2$, $N=50$ and $\beta$ in the
range $\beta=[0.625,0.626]$. As can be seen, the model is
compatible with the ACT data and the updated Planck constraint on
the tensor-to-scalar ratio.
However, we need to mention that this model leads to eternal inflation, at least classically, if no quantum corrections are taken into account.
Let us proceed to the non-local $R^2$ model, which we distinguish from pure power-law $F(R)$ gravity models due to the parametrization issues we mentioned earlier, see also Ref.~\cite{Odintsov:2025eiv}.

\subsection{Non-local Ghost-free $R^2$ Model}

\begin{figure}
\centering
\includegraphics[width=28pc]{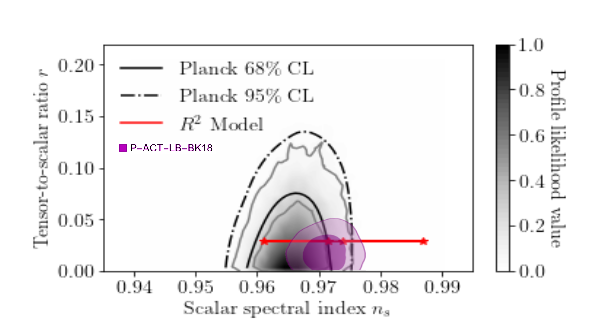}
\caption{Marginalized curves of the Planck 2018 data and the non-local $R^2$ gravity model, confronted with the ACT data, the Planck 2018 data, and the updated Planck constraints on the tensor-to-scalar ratio for $\alpha=0.064667699$, $N=60$ and $\beta$ in the range $\beta=[299,300.1]$. }\label{plot2}
\end{figure}

In the previous subsection, we studied the non-local power-law $F(R)$ gravity model, and in this section, we focus on the non-local ghost-free $R^2$ gravity model.
We discriminate due to the parameterization issues that the general power-law gravity is plagued with.
These issues were resolved in Ref.~\cite{Odintsov:2025eiv}, but unfortunately, the non-local ghost-free scenario is not easy to tackle analytically compared with the standard $F(R)$ gravity.
What we have in hand is an $F(R,\phi)$ theory, so the formalism of Ref.~\cite{Odintsov:2025eiv} does not apply, so we will adopt a formal approach with the standard parametrization.
In the case of non-local $R^2$ case, the $F(R)$ gravity function that appears in the action of Eq.~\eqref{modelnewnonlocalghostfreemodel} has the form,
\begin{align}
\label{powerlawfrsq}
F(R)=-\alpha R^2\, .
\end{align}
Now we shall assume that the scalar field obeys some constant-roll evolution \eqref{constantroll}, and in addition, we assume that the slow-roll conditions \eqref{slowrollhubblerate} hold for the inflationary evolution in general, thus we get for the scalar field evolution,
\begin{align}
\label{eqnmotionforscalarnewpowerlawsq}
3H(\beta+1)\dot{\phi}+\alpha R^2=0\, .
\end{align}
During inflation we approximately have $R\sim 12H^2$, thus we solve the equation \eqref{eqnmotionforscalarnewpowerlawsq} with respect to the derivative of the scalar field $\dot{\phi}$ and we get,
\begin{align}
\label{solutonfordotphi1}
\dot{\phi}\simeq -\frac{\gamma H^{3}}{\beta+1}\, ,
\end{align}
with $\gamma$ standing for,
\begin{align}
\label{gammadefinitiolocalB}
\gamma=\frac{12^2\alpha}{3}\, .
\end{align}
Upon substituting $\dot{\phi}$ from Eq.~\eqref{solutonfordotphi1} in Eq.~\eqref{firstfriedmanequation1}, and in addition, by recalling that during inflation we have,
\begin{align}
\label{approximaderivativefRsq}
f_R\sim 4 \phi \alpha R\, ,
\end{align}
we obtain $\phi$ expressed in terms of the Hubble rate,
\begin{align}
\label{resultingequationforphipowerlawsq}
\phi \sim A H^2 \, ,
\end{align}
with $A$ being defined as,
\begin{align}
\label{alphaparametersq}
A=\frac{\frac{12^4 \alpha^2}{2\cdot 9 (\beta +1)^2}+\frac{12^4 \alpha^2}{3}} {12^2 \alpha +12 \alpha }\, .
\end{align}
Using Eqs.~\eqref{resultingequationforphipowerlawsq} and \eqref{solutonfordotphi1}, we can express the slow-roll indices \eqref{slowrollparameters} in terms of the Hubble rate, in the following way,
\begin{align}
\label{slowrollparameterspowerlawmodelsq}
\epsilon_1=&\, \frac{\dot{H}}{H^2}\, , \quad \epsilon_2=3\beta\, ,
\quad \epsilon_3\simeq \frac{\dot{H}}{H^2}-\frac{\gamma }{2 A (\beta +1)} \, , \nonumber \\
\epsilon_4\simeq&\, -\frac{288 \alpha  A \beta  \Dot{H}^2}{\gamma (72 \alpha +A) H^4}-\frac{288 \alpha A \Dot{H}^2}{\gamma
(72 \alpha +A) H^4}+\frac{2 A \Dot{H}}{(72 \alpha +A) H^2}+\frac{144 \alpha  \Dot{H}}{(72 \alpha +A) H^2} \, ,
\end{align}
which can be rewritten in the following way,
\begin{align}
\label{slowrollparameterspowerlawmodellasttwoindicessq}
\epsilon_3\simeq -\frac{\gamma }{2 A (\beta +1)}+\epsilon_1 \, ,
\quad \epsilon_4\simeq -\frac{288 \alpha  A \beta
\epsilon_1^2}{\gamma  (72 \alpha +A)}-\frac{288 \alpha  A \epsilon_1^2}{\gamma  (72 \alpha +A)}
+\frac{2 A \epsilon_1}{72 \alpha +A}+\frac{144 \alpha \epsilon_1}{72 \alpha +A} \, .
\end{align}
Now we aim in finding in an approximate way the evolution during inflation quantified by the Hubble rate as a function of time, so by using the evolution for $\phi$ and $\dot{\phi}$ from Eqs.~\eqref{resultingequationforphipowerlawsq} and also \eqref{solutonfordotphi1} and also the differential equation \eqref{equationofmotionnew}, we get at leading order the following differential equation,
\begin{align}
\label{finalsemiapproximation1sq}
B H(t)^2-8\ 24 \alpha  \gamma H(t)^4 \Dot{H}+4\ 12 \alpha  \gamma H(t)^2\simeq 0\, ,
\end{align}
with $B$ in Eq.~\eqref{finalsemiapproximation1sq} being defined as,
\begin{align}
\label{bdefintion}
B=\frac{12^4 \alpha^2}{9}\, .
\end{align}
The differential equation of Eq.~\eqref{finalsemiapproximation1sq} has the analytic solution,
\begin{align}
\label{hubblesolutionsq}
H(t)=\frac{\sqrt[3]{192 \alpha \gamma \Lambda +B t +48 \alpha  \gamma  t}}{4 \sqrt[3]{\alpha } \sqrt[3]{\gamma }}\, ,
\end{align}
with $\Lambda$ being an integration constant.
This evolution deviates from the standard quasi-de Sitter evolution of $R^2$ gravity, since it describes a non-singular evolution if$\Lambda>0$, and a future Type III future singular evolution if $\Lambda<0$, see Ref.~\cite{Nojiri:2005sx}.
Now let us turn our focus on the phenomenology of the model, so by combining Eqs.~\eqref{slowrollparameterspowerlawmodellasttwoindicessq}) and \eqref{hubblesolutionsq}, and in addition, by solving the equation $\epsilon_1(t_f)=1$, we obtain the slow-roll parameters as functions of the $e$-foldings number and the free parameters of the model,
\begin{align}
\label{slowrollparameterspowerlawmodellasttwoindicessqnew}
\epsilon_1=&\, \frac{\alpha^2}{4 \alpha^2 N-1}\, , \quad
\epsilon_2=3\beta\, , \nonumber \\
\epsilon_3\simeq&\, \alpha^2 \left(\frac{1}{4 \alpha^2 N-1}-\frac{13 (\beta +1)}{4 \alpha^2
\left(6 \beta^2+12 \beta +7\right)}\right) \, , \nonumber \\
\epsilon_4\simeq&\, \frac{2 \alpha^2 A}{(72 \alpha +A) \left(4 \alpha^2 N-1\right)}
+\frac{144 \alpha^3}{(72 \alpha +A) \left(4 \alpha^2 N-1\right)} \nonumber \\
&\, -\frac{288 \alpha^5 A \beta }{\gamma  (72 \alpha +A) \left(4
\alpha^2 N-1\right)^2}
 - \frac{288 \alpha^5 A}{\gamma  (72 \alpha +A) \left(4 \alpha^2 N-1\right)^2} \, ,
\end{align}
which notice that are evaluated at the first horizon crossing time instance, and we made use of the $e$-foldings number relation $N=\int_{t_i}^{t_f}H\,\mathrm{d}t$.
Now the spectral index of the primordial scalar curvature perturbations takes the following form,
\begin{align}
\label{spectralindexofprimordialscalarcurvpert}
n_s=\frac{2 \left(-\frac{\gamma }{2 A \beta +2 A}+\frac{72 \alpha^3}{(72 \alpha +A) \left(4 \alpha^2 N-1\right)}
+\frac{A \left(288 \alpha^5 (\beta +1)-\alpha^2 \gamma +4 \alpha^4 \gamma N\right)}{\gamma  (72 \alpha +A) \left(1-4 \alpha^2 N\right)^2}-3
\beta \right)}{\frac{\alpha^2}{4 \alpha^2 N-1}+1}+1\, ,
\end{align}
and the tensor-to-scalar ratio takes the form,
\begin{align}
\label{tensortoscalarrsquaremodel}
r=\frac{52 (\beta +1)}{6 \beta^2+12 \beta +7}\, ,
\end{align}
hence it is $\beta$ dependent.
As it can be seen, the case $\beta=0$ results in a tensor-to-scalar ratio $r=7.42857$, which obviously indicates that the slow-roll case of the non-local $R^2$
$F(R)$ gravity model yields non-viable phenomenology.
Hence, we shall focus on the constant-roll case, which corresponds to $\beta\neq 0$.
To proceed with this model, we found various sets of values of the free parameters that may provide an ACT-compatible phenomenology, for example in this case, if we choose Planck units ($\kappa=1$) and take $\alpha=0.064667699$, $\beta=300.2$ and $N=60$ we get $n_s=0.9740$ and $r=0.028792$ which are well within the updated Planck constraint \eqref{planck} and the ACT constraint \eqref{act}.
In Fig.~\ref{plot2}, we present the confrontation of the non-local $R^2$ model with the Planck 2018 data and the updated ACT data, choosing $\alpha=0.064667699$, $N=60$, and $\beta$ in the range $\beta=[299,300.1]$.
As can be seen in this case too, the model is compatible with the ACT data and the updated Planck constraint on the tensor-to-scalar ratio.
Finally, let us consider the running of the spectral index for the
model at hand. The running of the spectral index is,
\begin{equation}\label{runningdef}
a_s=\frac{\mathrm{d} n_{\mathcal{S}}}{\mathrm{d} \ln k}\, ,
\end{equation}
where $k$ is the comoving wavenumber of a primordial mode. Then we
have,
\begin{equation}\label{runningdef1}
a_s=\frac{\mathrm{d} n_{\mathcal{S}}}{\mathrm{d} \ln
k}=\frac{\mathrm{d}
n_{\mathcal{S}}}{\mathrm{d}N}\frac{\mathrm{d}N}{\mathrm{d} \ln
k}\, ,
\end{equation}
where $N$ is the $e$-foldings number. Also, by using
$\frac{\mathrm{d}N}{\mathrm{d} \ln k}=\frac{1}{1-\epsilon_1}$, we
get $a_s$, as follows,
\begin{equation}\label{runningdefmainfinal}
a_s=\frac{1}{1-\epsilon_1}\frac{\mathrm{d}
n_{\mathcal{S}}}{\mathrm{d}N}\, .
\end{equation}
For the model we consider here and for the values of the free
parameters considered earlier, we obtain $a_s=0.0000238$. Note
that the ACT constraints indicate that $a_s=[-0.0042,0.0166]$ at
$95\%$ confidence level. Note however that the ACT data also
indicate a positive running of the spectral index, but the result
is at $1.2\sigma$ conflict with the $68\%$ constraints, thus it is
insignificant. Nevertheless the prospect of having a positive
running of the spectral index is something reportable and should
further be studied, because if confirmed at a higher confidence
level, this would eliminate many theories from being viable
inflationary theories. We aim to report on this soon. Also note
that the model of the previous section yields a zero running of
the spectral index because it is a power-law evolution.

\section{Conclusions}

In this article, we reviewed and analyzed the ghost-free non-local
$F(R)$ gravity framework. We demonstrated how a ghost-free
non-local theory of $F(R)$ gravity can be produced, and also we
showed that a de Sitter solution exists for that framework. We
also showed that ghost-free non-local $F(R)$ gravity can be cast
in the form of $F(R,\phi)$ theory of gravity. We chose two
distinct models of $F(R)$ gravity, a non-local power-law $F(R)$
gravity and a non-local $R^2$ gravity model, and we confronted
these models with the ACT constraints on the spectral index of the
scalar perturbations and the Planck/BICEP updated constraints on
the tensor-to-scalar ratio. As we showed, both these models can be
well accommodated within the ACT and Planck/BICEP  constraints,
thus they can provide a viable inflationary phenomenology.
Non-local models of gravity are of great importance, since these
can also emulate dark matter in some
contexts~\cite{Deffayet:2025lwl, Deffayet:2024ciu}. With this
work, we demonstrated that the ghost-free non-local version of
$F(R)$ gravity can lead to viable inflation, which is compatible
with the ACT and updated BICEP/Planck constraints. About the
future prospects of this work, the first aspect that needs proper
discussion is the fact that the present model yields a slightly
positive running, in contrast to pure $F(R)$ gravity models and
scalar-tensor models. This behavior is not accidental and we came
across the same behavior in other $F(R,\phi)$ models. So one may
generalize the current work and study further $F(R,\phi)$
generalizations and also focus on the running of the spectral
index. We aim to report on this soon.

\end{document}